\documentclass[aps,prd,reprint,amsmath,amssymb,nofootinbib,floatfix,superscriptaddress]{revtex4-2}
\usepackage{graphicx}
\usepackage{dcolumn}
\usepackage{multirow}
\usepackage[utf8]{inputenc}
\usepackage{xcolor}
\usepackage[unicode=true,colorlinks,linkcolor=blue,anchorcolor=blue,urlcolor=blue,citecolor=blue,breaklinks=true]{hyperref}
\usepackage{epstopdf}

\newcommand{\ii}{\mathrm{i}}

\newcommand{\pararrow}{\mathord{\buildrel{\lower3pt\hbox{$\scriptscriptstyle\leftrightarrow$}}\over {\partial}}} 
\newcommand{\pararrowk}[1]{\mathord{\buildrel{\lower3pt\hbox{$\scriptscriptstyle\leftrightarrow$}}\over {\partial}\hspace*{-0.18em}{}^#1}\hspace*{-0.18em} \,} 
\newcommand{\mymel}[3]{\langle #1| #2 | #3\rangle}
\usepackage{bm}
\usepackage{slashed}

\newcommand{\pcs}{\textit{P}_{\bar{c}s}^{\,-}}
\newcommand{\pcsstar}{\textit{P}_{\bar{c}s}^{\,\ast-}}
\newcommand{\lmdb}{\Lambda_b^0}

\newcommand{\qfnu}{\affiliation{College of Physics and Engineering, Qufu Normal University, Qufu 273165, China}}
\newcommand{\tju}{\affiliation{Center for Joint Quantum Studies and Department of Physics, School of Science,\\ Tianjin University, Tianjin 300350, China}}

\begin{document}
	
	\title{Production and decay of anticharmed pentaquark state with quark content \texorpdfstring{$\bar{c}sudd$}{}}
	\author{Shi-Dong Liu} \tju\qfnu
	\author{Xiao-Yun Wang} \tju
	\author{Xiao-Hai Liu}\email{xiaohai.liu@tju.edu.cn}\tju
	\author{Gang Li}\email{gli@qfnu.edu.cn} \qfnu

\begin{abstract}
We use an effective Lagrangian approach to investigate the production of the anticharmed pentaquark state $P_{\bar{c}s}^{(*)-}$ with the minimal quark content $\bar{c}sudd$ in the $\Lambda_b^0$ decay. In our calculation, the $P_{\bar{c}s}^{(*)-}$ is considered as the molecule state of the $nD_{s0(s1)}^{-}$ in an $S$ wave, and its production and decay occur via triangle loops at the hadron level. With the moderate model parameters, the computed results suggest that the branching fractions of $\Lambda_b^0\to P_{\bar{c}s}^{(*)-} D^+ $ can reach approximately $10^{-4}$. The possible decay modes $ P_{\bar{c}s}^{(*)-} \to \Lambda D^{(\ast)-}$ and $ P_{\bar{c}s}^{(*)-} \to \Sigma D^{(\ast)-}$ are also investigated. It is hoped that these predictions could be helpful in searching for the anticharmed-strange pentaquark candidates in future LHCb experiments.
\end{abstract}
\date{\today}
	
\maketitle
\section{Introduction}\label{sec:intro}
Within the traditional quark model, mesons are made of a quark-antiquark pair, while baryons are composed of three quarks. Although the quark model allows the existence of other complicated quark compositions \cite{gell-mann1964PL8-214,Zweig:1964ruk}, the simplest cases of the meson and baryon should be the norm. The multiquark states that contain more than three quarks usually have higher masses, especially when among them there are heavy quarks. Consequently, earlier studies of the multiquark states focus mainly on those made of light quarks, i.e., the $u$, $d$, and $s$ quarks (see reviews \cite{liu2019PPNP107-237,chen2016PR639-1,guo2018RMP90-015004,karliner2018ARNPS68-17} and references therein).

The situation of multiquark states changes significantly in 2003 when the $X(3872)$ was observed by the Belle Collaboration \cite{choi2003PRL91-262001}. Nowadays, the $X(3872)$ is the most well-studied tetraquark state. With development of the experimental techniques, lots of tetraquark-state candidates have been observed in various experiments \cite{workman2022PTEP2022-083C01}, which are usually dubbed by the $XYZ$ states, e.g., the $Z_c(3900)$ \cite{ablikim2013PRL110-252001,liu2013PRL110-252002,xiao2013PLB727-366,abazov2018PRD98-052010} and $Z_b(10610/10650)$ \cite{bondar2012PRL108-122001,adachi2012a[x-,krokovny2013PRD88-052016,garmash2015PRD91-072003}. Such tetraquark states contain a heavy quark and antiquark pair ($c\bar{c}$ or $b\bar{b}$). In 2022, the LHCb Collaboration reported a double-charm tetraquark state candidate $T_{cc}^+$, whose minimal quark content is $cc\bar{u}\bar{d}$ \cite{aaij2022NP18-751,aaij2022NC13-3351}. The experimental observation of this double-charm tetraquark candidate convinces us of the existence of the possible $bc\bar{u}\bar{d}$ and $bb\bar{u}\bar{d}$ tetraquark states. Various theoretical interpretations, e.g., the compact multiquark, hadronic molecule, cusp effect, and hadro-quarkonium, can be found in the reviews \cite{chen2016PR639-1,dong2017PPNP94-282,ali2017PPNP97-123,esposito2017PR668-1,lebed2017PPNP93-143,olsen2018RMP90-015003,guo2018RMP90-015004,yuan2018IJMPA33-1830018,karliner2018ARNPS68-17,liu2019PPNP107-237,brambilla2020PR873-1,chen2022RPP86-026201,meng2023PR1019-2266} and references therein.
	
The pentaquark state candidate was first observed experimentally in 2015 when the LHCb Collaboration announced two structures in the $J/\psi p$ invariant mass distribution in $\Lambda_b^0\to J/\psi K^- p$ decay process \cite{aaij2015PRL115-072001}. This two structures have minimal quark content $c\bar{c}uud$ and masses close to $4.4~\mathrm{GeV}$, called $P_c(4380)^+$ and $P_c(4450)^+$. Four years later, refined LHCb experiments discovered with higher statistical significance a new narrow pentaquark candidate $P_c(4312)^+$ and two sub-structures, $P_c(4440)^+$ and $P_c(4457)^+$, in the early reported $P_c(4450)^+$ \cite{aaij2019PRL122-222001}, while the $P_c(4380)$ appears to be observed again \cite{meng2023PR1019-2266}. Later, a strange counterpart $P_{cs}(4459)^0$ with minimal quark content $c\bar{c}sud $ was found by the LHCb Collaboration in the $J/\psi \Lambda$ invariant mass distribution of the decay $\Xi_{b}\to J/\psi \Lambda K^-$ \cite{aaij2021SB66-1391}. Apart from in the decay  $\Lambda_b^0\to J/\psi K^- p$, the LHCb, using the data collected in proton-proton collisions between 2011 and 2018, found another new pentaquark resonance $P_c(4337)^+$ in the $B_s^0\to J/\psi p\bar{p}$ \cite{aaij2022PRL128-062001}. These pentaquark structures have widths between 6 and 30 MeV, whereas their quantum numbers are not determined yet and have various possibilities, depending on the employed interpretations \cite{aaij2019PRL122-222001,aaij2022PRL128-062001,meng2023PR1019-2266}. Moreover, they are all charmonium-pentaquark states. In order to interpret the nature of these $P_{c(s)}$'s, different scenarios have been proposed, such as the kinematic effect \cite{guo2015PRD92-071502,liu2016PLB757-231,bayar2016PRD94-074039}, molecular picture \cite{chen2019PRD100-011502,xiao2019PRD100-014021,liu2019PRL122-242001,guo2019PLB793-144,meng2019PRD100-014031,chen2019PRD100-051501,burns2019PRD100-114033,gutsche2019PRD100-094031,wang2019J11-108,voloshin2019PRD100-034020,wu2019PRD100-114002,lin2019PRD100-056005,yamaguchi2020PRD101-091502,du2020PRL124-072001,wang2020PRD102-036012,xiao2020PRD102-056018,dong2021PP41-65,peng2022PRD105-034028,kuang2020EPJC80-433}, threshold cusp \cite{kuang2020EPJC80-433,nakamura2021PRD104-L091503}, and compact pentaquark model \cite{Giron:2019bcs,ali2019PLB793-365,zhu2019PLB797-134869,stancu2019EPJC79-957,wang2020IJMPA35-2050003,giron2021PRD104-114028}. Actually, the charmonium-pentaquark states have been theoretically predicted already in 2010 within the molecule framework \cite{wu2010PRL105-232001}.

In the heavy flaovr sector, besides the well studied two-body hadronic molecules, some few-body systems have also been studied in literature, such as the hidden charmed systems $D\bar{D}K$~\cite{Wu:2020job}, $D\bar{D}^*K$~ \cite{Wu:2020job,Ma:2017ery,Ren:2018pcd},  $ D\bar{D} \rho$~\cite{Durkaya:2015wra}, $\Xi_{cc}\bar{\Xi}_{cc}\bar{K}$~\cite{Wu:2020rdg}, and $\Sigma_c \bar{D}\bar{K}$~\cite{Wu:2021gyn}, the singly charmed systems $DNN$~\cite{Bayar:2012dd}, $DK\bar{K}$~\cite{MartinezTorres:2012jr,Debastiani:2017vhv}, and $KDN$($\bar{K}DN$, $\bar{K}\bar{D}N$)~\cite{Xiao:2011rc,Yamagata-Sekihara:2018gah}, the doubly charmed systems $DDK$~\cite{MartinezTorres:2018zbl,Wu:2019vsy,Huang:2019qmw,Pang:2020pkl}, $BDD$~\cite{Dias:2017miz}, and $D^{(*)}D^*K$~\cite{Ma:2017ery,Ren:2024mjh}, the triply charmed four-body system $DDDK$~\cite{Wu:2019vsy}, the quadruply charmed system $\Xi_{cc}\Xi_{cc}\bar{K}$~\cite{Wu:2020rdg}, and so on. See Ref.~\cite{MartinezTorres:2020hus} for a review about the few-body systems. Very recently, the analysis of the invariant-mass distributions of the $D^-\Lambda$ in the decay $\Lambda_b^0\to D^+D^-\Lambda$ conducted in the LHCb experiments \cite{aaij2024a[x-} implies existence of suspected anticharmed pentaquark states with masses around $3300~\mathrm{MeV}$ and $3650~\mathrm{MeV}$, whose minimal quark content is the $\bar{c} sudd$, denoted as $\pcs$ hereafter. This experimental observation to some extent is consistent with the prediction of the $\bar{K}\bar{D}N$ three-body molecule given in Ref. \cite{Yamagata-Sekihara:2018gah}. For the $\bar{K}\bar{D}N$ system, the interactions for every pair of two hadrons are supposed to be attractive. The $D_{s0}^\ast(2317)$ and $\Lambda(1405)$ are widely believed to be the $\bar{K}\bar{D}$ and $\bar{K}N$ bound states, respectively, especially within the framework of unitarized chiral perturbation theory~\cite{Oset:1997it, Oller:2000fj, Oset:2001cn, guo2018RMP90-015004,Guo:2006fu,Guo:2006rp,Guo:2009ct,Kolomeitsev:2003ac,Altenbuchinger:2013vwa, Guo:2015dha,Liu:2012zya,Yao:2015qia}. For the $\bar{K}\bar{D}$ or $\bar{K}N$ interaction, the Weinberg-Tomozawa term gives the most important contribution at low energies, and the $\bar{K}$ acts as the glue that binds the $\bar{K}\bar{D}N$ system. Although the $\bar{D}N$ interaction in the isoscalar channel is weak, it is still attractive due to the coupled channel interactions~\cite{Gamermann:2010zz}. Therefore, it is highly expected that the  $\bar{K}\bar{D}N$ three-body bound state may exist, or in another way, the $ND_{s0}^\ast(2317)$ and $\bar{D}\Lambda(1405)$ molecular states may exist. Likewise, the $D_{s1}(2460)$ is widely supposed to be the $\bar{K}\bar{D}^*$ molecular state, therefore the existence of $\bar{K}\bar{D}^*N$ molecular state is also expected. The eigenenergy of $\bar{K}\bar{D}N$ system is estimated to be about $(3244- 17\ii)$ MeV in Ref. \cite{Yamagata-Sekihara:2018gah}. This three-body molecular state may correspond to the resonance-like structure around 3300 MeV observed in $D^-\Lambda$ spectrum \cite{aaij2024a[x-}.

In this work, we assume that the experimentally observed process $\Lambda_b^0\to D^+D^-\Lambda$ by LHCb \cite{aaij2024a[x-} could occur via a cascade process, namely, $\Lambda_b^0\to D^+\pcs \to D^+D^-\Lambda$. It is further assumed that the $\pcs$ is the molecular state of the $nD_{s0}^{\ast -}(2317)$. According to the heavy quark spin symmetry, there would exist the partner, $\pcsstar$, as the $nD_{s1}^-(2460)$ molecule\footnote{For simplicity, we shall, respectively, use $D_{s0}^-$ and $D_{s1}^-$ to represent the $D_{s0}^{\ast -}(2317)$ and $D_{s1}^-(2460)$, unless otherwise stated.}. We predict the productions and decays of the $\pcs$ and $\pcsstar$ using an effective Lagrangian approach at the hadronic level. In the following, we shall first give the theoretical consideration in Sec. \ref{sec:lags}, then in Sec. \ref{sec:results} present the calculated results and discussion, and finally give a summary in Sec. \ref{sec:summary}.

\section{Theoretical Consideration}\label{sec:lags}
	\begin{figure}
		\centering
		\includegraphics[width=0.98\linewidth]{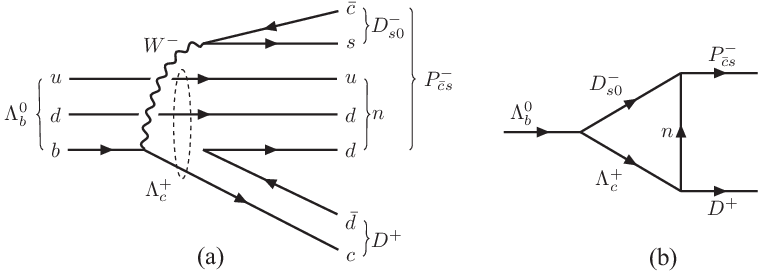}
		\caption{The Feynman diagram for the production of the $P_{\bar{c}s}^-$ at the quark (a) and hadron (b) levels. The $\pcsstar$ production process can occur via the triangle loop where the $D_{s0}^-$ is replaced by the $D_{s1}^-$.}
		\label{fig:feyndiagsproduction}
	\end{figure}
	
At the quark level, the $\lmdb$ indirectly decays into the $\pcs$ and $D^+$ via the schematic diagram shown in Fig. \ref{fig:feyndiagsproduction}(a): the $b$ quark converts into the $c$ quark, with emission of a $W^-$ boson; Subsequently, the $W^-$ boson splits into the $\bar{c}s$ quark pair, which momentarily form the $D_{s0}^-$ meson; in the meantime, a pair of $d\bar{d}$ quarks are created from the vacuum, the $d$ of which and the spectator quarks $d$ and $u$ combine together to make the neutron, while the other $\bar{d}$ and the $c$ quark produced in weak process form the $D^+$ meson; the $D_{s0}^-$ and neutron are bounded into the $\pcs$. The processes at quark level mentioned above could be described at the hadronic level by the triangle loop in Fig. \ref{fig:feyndiagsproduction}(b): The $\lmdb$ decays firstly into the $\Lambda_c^+$ and $D_{s0}^-$, then the two particles couple further to the final particle by exchanging the neutron.
	
\begin{figure}
	\centering
	\includegraphics[width=0.80\linewidth]{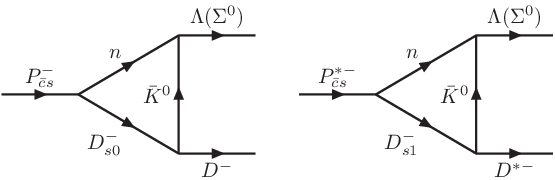}
	\caption{The triangle loops contributing to the $\pcs\to\Lambda(\Sigma^0) D^-$ (left) and $\pcsstar\to\Lambda(\Sigma^0) D^{\ast -}$ (right). The diagrams for the $P_{\bar{c}s}^{(\ast)-}\to\Sigma^- \bar{D}^{(\ast)0}$ can be obtained by replacing the meson $\bar{K}^0$ with the $K^-$.}
	\label{fig:feyndiagsdecay}
\end{figure}

If the molecule state $\pcs$ as the $nD_{s0}^-$ exists, the $nD_{s1}^-$ molecular state, denoted as $\pcsstar$, can also be likely to be formed based on the heavy quark spin symmetry. As the $nD_{s0(s1)}$ molecule, the $P_{\bar{c}s}^{(\ast)-}$ would decay with relatively high rate into its components $n$ and $D_{s0(s1)}^- $. Hence, we describe the decay process $P_{\bar{c}s}^{(\ast)-}\to\Lambda D^{(\ast)-}$ via the triangle loops shown in Fig. \ref{fig:feyndiagsdecay}. According to the $SU(3)$ symmetry, the processes $P_{\bar{c}s}^{(\ast)-}\to\Sigma^0 D^{(\ast)-}$ and $P_{\bar{c}s}^{(\ast)-}\to\Sigma^- \bar{D}^{(\ast)0}$ are expected to also occur.
	
\subsection{The effective Lagrangians}
	
In terms of the factorization ansatz, the amplitudes of the decay $\Lambda_b^0\to \Lambda_c^+D_{s0}^-$ can be described as \cite{liu2024PRD109-056014,bauer1987ZPC34-103}
\begin{align}
	\mathcal{M} &= \frac{G_{F}}{\sqrt{2}}V_{cb}V_{cs}a_1\mymel{D_{s0}^-}{\bar{s}\gamma^\mu(1-\gamma_5)c}{0}\nonumber\\
	&\times \mymel{\Lambda_c^+}{\bar{c}\gamma_\mu (1-\gamma_5) b}{\Lambda_b^0}\,,
\end{align}
where $G_F$ is the Fermi coupling constant \cite{vanritbergen1999PRL82-488}, $V_{cb}$ and $V_{cs}$ are the Cabibbo-Kobayashi-Maskawa (CKM) matrix elements, and $a_1=0.88$ is the effective Wilson coefficient \cite{liu2024PRD109-056014}. Furthermore, the current matrix element $\mymel{\Lambda_c^+}{\bar{c}\gamma_\mu (1-\gamma_5) b}{\Lambda_b^0}$, responsible for the transition of the $b$ to $c$, reads
\begin{align}\label{eq:LamcLamb}
   &\mymel{\Lambda_c^+}{\bar{c}\gamma_\mu (1-\gamma_5) b}{\Lambda_b^0}=\bar{u}(p') \big[f_1^V(q^2)\gamma_\mu -f_2^V(q^2) \frac{\ii\sigma_{\mu\nu}q^\nu}{m} \nonumber\\
	&+ f_3^V(q^2)\frac{q_\mu}{m} -\big(f_1^A(q^2)\gamma_\mu -f_2^A(q^2) \frac{\ii\sigma_{\mu\nu}q^\nu}{m}\nonumber\\
	&+f_3^A(q^2)\frac{q_\mu}{m}\big)\gamma_5\big]u(p) \equiv \bar{u}(p')X_\mu u(p),
\end{align}
where $\sigma_{\mu\nu} = \ii [\gamma_\mu, \gamma_\nu]/2$ and $q=p-p'$ with the $p^\prime$ and $p$ being the four-momentum of the $\Lambda_c^+$ and $\Lambda_b^0$, respectively. The $f_i^{V(A)}(q^2)$ are the form factors for the transition $\Lambda_b^0\to\Lambda_c$ and in present calculations we adopt the following form\footnote{It is noted that the Ref. \cite{gutsche2015PRD91-074001} used the different form: $f_i^{V(A)}(q^2) = F_i^{V(A)}(0)/(1-a s+b s^2)$ with $s=q^2/m^2$. These two expressions both exhibit high accuracy (see Appendix \ref{app:fff}). }
\begin{equation}\label{eq:formfactors}
	f_i(q^2) = F_i(0) \left(\frac{\Lambda_i^2}{q^2-\Lambda_i^2}\right)^2\,.
\end{equation}
Here the parameters $F_i(0)$ and $\Lambda_i$ can be determined by fitting Eq. \eqref{eq:formfactors} through the numerical data obtained within the covariant confined quark model \cite{gutsche2015PRD91-074001}, summarized in Table \ref{tab:formfactorvalue}. The current matrix element $\mymel{D_{s0}^-}{\bar{s}\gamma^\mu(1-\gamma_5)c}{0}$, describing the creation of the $D_{s0}^-$ from the vacuum, is given as \cite{cheng2004PRD69-074025,liu2024PRD109-056014}
\begin{equation}
	\mymel{D_{s0}^-}{\bar{s}\gamma^\mu(1-\gamma_5)c}{0}=f_{D_{s0}^-}q^\mu\,,
\end{equation}
while the element $\mymel{D_{s1}^-}{(\bar{s}c)}{0}$ is in the following form
\begin{equation}
	\mymel{D_{s1}^-}{\bar{s}\gamma^\mu(1-\gamma_5)c}{0} = m_{D_{s1}}f_{D_{s1}^-}\epsilon^{\ast \mu}
\end{equation}
where $f_{D_{s0}^-}=114~\mathrm{MeV}$ and $f_{D_{s1}^-} = 194~\mathrm{MeV}$ are, respectively, the decay constants of the $D_{s0}^-$ and $D_{s1}^-$, in terms of the lattice QCD calculation \cite{bali2017PRD96-074501}. Alternatively, $f_{D_{s0}^-}=333~\mathrm{MeV}$ and $f_{D_{s1}^-} = 345~\mathrm{MeV}$, obtained using the QCD sum rule \cite{wang2015EPJC75-427}. It is noted that the values of the $f_{D_{s0}^-} $ and $f_{D_{s1}^-} $ change by about an order of magnitude for different interpretations of the $D_{s0(s1)}^-$ \cite{liu2024PRD109-056014,faessler2007PRD76-014003}.

	\begin{table}
		\caption{The values of the parameters $F_i^{V(A)}(0)$ and $\Lambda_i^{V(A)}$ for the transition form factors $f_i^{V(A)}$. They are obtained by fitting Eq. \eqref{eq:formfactors} through the numerical data shown in the Fig. 2 of Ref \cite{gutsche2015PRD91-074001}. The $\Lambda_i^{V(A)}$ are in $\mathrm{GeV}$.}
		\label{tab:formfactorvalue}
		\begin{ruledtabular}
			\begin{tabular}{lccc}
				$i$&$1$	&$2$& $3$\\
				\colrule
				$F_i^{V}$&$0.550$ & $0.110$ & $-0.023$\\
				$F_i^{A}$&$0.543$ & $0.018$ & $-0.124$\\
				$\Lambda_i^{V}$&$6.627$	&$6.222$& $7.215$\\
				$\Lambda_i^{A}$&$6.662$	&$8.359$& $6.160$\\
			\end{tabular}
		\end{ruledtabular}
	\end{table}

Within the molecular framework, we consider the $\pcs$ as a molecular state of the $nD_{s0}^-$ in an $S$ wave, thereby having the quantum numbers $J^P=1/2^+$. The coupling of the $\pcs$ to its components can be constructed as \cite{zou2003PRC67-015204}
\begin{equation}\label{eq:Lpcs}
	\mathcal{L}_{\pcs}=g_1\bar{n}P_{\bar{c}s}D_{s0}\,.
\end{equation}
When we consider the partner $\pcsstar$ as the molecular state of the $nD_{s1}^-$ in an $S$ wave, the quantum numbers of the $\pcsstar$ could be $J^P=1/2^+$ or $J^P=3/2^+$. The interactions between the $\pcsstar$ and its components read \cite{zou2003PRC67-015204,wu2021CPL38-071301}
\begin{subequations}\label{eq:Lpcss}
	\begin{align}
		\mathcal{L}_{\pcsstar}^{1/2}&= g_1' \bar{n} \gamma_5(g_{\mu\nu}-\frac{p_\mu p_\nu}{m^2})\gamma^\nu P_{\bar{c}s}^\ast D_{s1}^\mu\,,\\
		\mathcal{L}_{\pcsstar}^{3/2} &= g_1''\bar{n} P_{\bar{c}s\mu}^\ast D_{s1}^\mu\,.
	\end{align}
\end{subequations}

If the $\pcs$ and $\pcsstar$ are bound states, the coupling constants $g_1$'s in Eqs. \eqref{eq:Lpcs} and \eqref{eq:Lpcss} could be estimated by employing the compositeness relation~\cite{Weinberg:1962hj,lin2017PRD95-114017,wu2021CPL38-071301}
\begin{equation}\label{eq:couplingg1}
	g_1=\left( \frac{\pi}{m_P m_n}\frac{(m_{D_{s0(s1)}}+m_n)^{5/2}}{\sqrt{m_{D_{s0(s1)}}m_n}}\sqrt{32 E_b}\right)^{1/2},
\end{equation}
where $E_b = m_n+m_{D_{s0(s1)}}-m_P$ with $m_P,\,m_n$, and $m_{D_{s0(s1)}}$ being the masses of the $P_{\bar{c}s}^{(\ast)-}$, $n$, and $D_{s0(s1)}^-$, respectively. On the other hand, if $m_{\pcs}$ is greater than the $nD_{s0}^-$ or $D^-\Lambda(1405)$ threshold, we will employ the approach proposed in Ref.~\cite{Meissner:2015mza} to determine the effective coupling, where the compositeness relation was extended to the resonances. Following Ref.~\cite{Meissner:2015mza} and imposing saturation of the compositeness relation and the width of $\pcs$ by two channels, $nD_{s0}^-$ (channel 1) and $D^-\Lambda(1405)$ (channel 2), we can determine the effective couplings by solving the following equations
\begin{eqnarray}
 && |\gamma_1|^2 \left| \frac{\partial G_1 (s_{\pcs})}{\partial s} \right| +|\gamma_2|^2 \left| \frac{\partial G_2 (s_{\pcs})}{\partial s} \right| = 1, \label{eq:comp1}\\
 && \frac{q_1 |\gamma_1|^2}{8\pi m_{\pcs}^2}  +\frac{q_2 |\gamma_2|^2}{8\pi m_{\pcs}^2} = \Gamma_{\pcs}, \label{eq:comp2}
\end{eqnarray}
where $G_{1,2}(s)$ is the scalar unitary loop function~\cite{Oller:2000fj,Oller:2006jw}, $q_{1,2}$ is the decay momentum, and $\gamma_{1,2}$ is the coupling of the $\pcs$ to pertinent channel. The effective coupling in Eq.~(\ref{eq:couplingg1}) can be expressed in terms of $\gamma_1$, i.e., $|g_1|=|\gamma_1|/(4m_{\pcs} m_n)^{1/2}$. The $s_{\pcs}$ in Eq.~(\ref{eq:comp1}) corresponds to the pole position of $\pcs$ in the complex energy plane and takes the form  
\begin{equation}
    s_{\pcs}=\left( m_{\pcs} -i\frac{\Gamma_{\pcs}}{2} \right)^2.
\end{equation}
The effective couplings for $\pcsstar$ are estimated similarly. See Refs.~\cite{Baru:2003qq,Hyodo:2011qc,Aceti:2012dd,Sekihara:2014kya,Guo:2015daa,Kang:2016ezb} for more discussions about the compositeness relation for resonances.

The interaction of the $\Lambda_c^+D^+n$ can be described by the Lagrangian in the form \cite{khodjamirian2011J09-106,qian2022PLB833-137292}
\begin{equation}
	\mathcal{L}_{\Lambda_cnD}=\ii g_2 \bar{\Lambda}_c \gamma_5 D N\,.
\end{equation}
According to the QCD light-cone sum rule \cite{khodjamirian2011J09-106}, the coupling constant $g_2$ is estimated to be about $13.8$.

The effective Lagrangian for the $\Lambda(\Sigma) n \bar{K}$ could be described as \cite{Jackson:2015dva,Liu:2008ck}
\begin{equation}
	\mathcal{L}_{\Lambda n\bar{K}} =-\ii g_3 \bar{n}\gamma_5 \Lambda(\Sigma) K\,.
\end{equation}
The couplings for the $\Lambda$ and $\Sigma$ are related to each other by the flavor $SU(3)$ symmetry, and in the following calculations we take $g_3=-13.24$ for the $\Lambda$ and $g_3=3.58$ for the $\Sigma^{0(-)}$ \cite{Jackson:2015dva,Liu:2008ck}.
From the point of view of the heavy quark effective theory and the chiral symmetry, the interaction for the $D_{s0(s1)}^-D^-\bar{K}^0$ read \cite{casalbuoni1997PR281-145}
\begin{equation}
	\mathcal{L}_{S} = \frac{2h}{f_\pi}\partial_\mu K v^\mu \left( \bar{D}_{s0}D - \bar{D}_{s1b}^\nu D^{\ast}_\nu \right)
\end{equation}
Here $v=(1,\bm{0})$ is the velocity of the heavy mesons, $f_\pi=132~\mathrm{MeV}$ is the pion decay constant, and the coupling $h = -0.56$ \cite{casalbuoni1997PR281-145}.
\subsection{The decay amplitudes}

According to the above effective Lagrangian, the amplitude for the process $\Lambda_b^0 (p)\to [\Lambda_c^+(p')\,D_{s0}^-(q)\,n(l)]\to \pcs(k)\, D^+(p-k)$ where the $\pcs$ is considered as the $nD_{s0}^-$ molecule, is
\begin{align}
	\mathcal{M} &= \ii^3 \int \frac{\mathrm{d}^4l}{(2\pi)^4}\big[g_1 \bar{u}(k)\big](\slashed{l}+m_n)\big[\ii g_2 \gamma_5\big](\slashed{p'}+m_{\Lambda_c})\nonumber\\
	&\times \big[\frac{G_{F}}{\sqrt{2}}V_{cb}V_{cs}a_1f_{D_{s0}^-}q^\mu X_\mu u(p)\big] \frac{1}{{p'}^2-m_{\Lambda_c}^2}\nonumber\\
	&\times \frac{1}{q^2-m_{D_{s0}}^2}\frac{1}{l^2-m_n^2}\mathcal{F}(l^2,m_n^2)\,.
\end{align}

The vector structure $X_\mu$ is described explicitly in Eq. \eqref{eq:LamcLamb}.
The form factor $\mathcal{F}(l^2,m_n^2)$ is introduced to describe the structure and off-shell effects of the exchanged particle, in the monopole form
\begin{equation}
	\mathcal{F}(l^2,m_n^2) = \frac{m_n^2 - \Lambda_E^2}{l^2 - \Lambda_E^2}\,,
\end{equation}
where $\Lambda_E = m_n + \alpha \Lambda_{\mathrm{QCD}}$ with $\Lambda_{\mathrm{QCD}} = 220~\mathrm{MeV}$. The model parameter $\alpha$ is usually regarded as an undetermined parameter in the vicinity of unity. In present calculations, we take the $\alpha$ from $1.0$ to $2.0$.

Similarly, we could get the amplitudes for the process $\Lambda_b^0 (p)\to [\Lambda_c^+(p')\,D_{s1}^-(q)\,n(l)]\to \pcsstar(k) D^+(p-k)$ in which the $\pcsstar$ is interpreted as the $nD_{s1}^-$ molecular state. In the case of the $1/2^+$ $\pcsstar$, the amplitude reads
\begin{align}
	\mathcal{M} &= \ii^3 \int \frac{\mathrm{d}^4l}{(2\pi)^4}\big[g_1' \bar{u}(k) \gamma^\alpha\gamma_5\tilde{g}_{\alpha\beta}(k,m_P)\big](\slashed{l}+m_n)\nonumber\\
	&\times \big[\ii g_2 \gamma_5\big](\slashed{p'}+m_{\Lambda_c}) \big[\frac{G_{F}}{\sqrt{2}}V_{cb}V_{cs}a_1f_{D_{s1}^-}m_{D_{s1}} X_\mu u(p)\big] \nonumber\\
	&\times \frac{1}{{p'}^2-m_{\Lambda_c}^2}\frac{\tilde{g}^{\mu\beta}(q,m_{D_{s1}})}{q^2-m_{D_{s1}}^2}\frac{1}{l^2-m_n^2}\mathcal{F}(l^2,m_n^2)\,,
\end{align}
whereas the amplitude for the $3/2^+$ $\pcsstar$ is
\begin{align}
	\mathcal{M} &= \ii^3 \int \frac{\mathrm{d}^4l}{(2\pi)^4}\big[g_1'' \bar{u}_\alpha(k)\big](\slashed{l}+m_n)\big[\ii g_2 \gamma_5\big]\nonumber\\
	&\times (\slashed{p'}+m_{\Lambda_c}) \big[\frac{G_{F}}{\sqrt{2}}V_{cb}V_{cs}a_1f_{D_{s1}^-}m_{D_{s1}} X_\mu u(p)\big] \nonumber\\
	&\times \frac{1}{{p'}^2-m_{\Lambda_c}^2}\frac{\tilde{g}^{\mu\alpha}(q,m_{D_{s1}})}{q^2-m_{D_{s1}}^2}\frac{1}{l^2-m_n^2}\mathcal{F}(l^2,m_n^2)\,.
\end{align}
Here $\tilde{g}_{\alpha\beta}(p,m) \equiv -g_{\alpha\beta} + p_\alpha p_\beta / m^2$.

The partial width of the $\Lambda_b^0\to P_{\bar{c}s}^{(\ast)-} D^+$ is given by 
\begin{equation}
	\Gamma = \frac{1}{2} \frac{1}{8\pi} \frac{\left| \bm{k}\right| }{m_{\Lambda_b^0}^2} \sum_{\mathrm{spin}} \left| \mathcal{M}\right|^2\,,
\end{equation}
where the factor $1/2$ is due to the spin average over the initial particle $\Lambda_b^0$, $\bm{k}$ is the momentum of the $P_{\bar{c}s}^{(\ast)-}$ or the $D^+$ in the rest frame of the $\Lambda_b^0$, and $\sum_{\mathrm{spin}}$ means the summation over the spins of the initial and final states. For the spin-$3/2$ state, its vector-spinor satisfies \cite{chung1971x-}
\begin{equation}
	\sum_s u_\mu(p,s)\bar{u}_\nu(p,s) =(\slashed{p}+m)\big[\tilde{g}_{{\mu\nu}}+\frac{1}{3}\tilde{g}_{\mu\alpha}\gamma^\alpha\tilde{g}_{\nu\beta}\gamma^\beta\big].
\end{equation}

By the same token, we can obtain three amplitudes for the $P_{\bar{c}s}^{(\ast)-}$ decays, of which the diagrams are shown in Fig. \ref{fig:feyndiagsdecay}. For the case of the $\pcs$ as the $nD_{s0}^-$ molecular state, the amplitude for the $\pcs(p) \to [n(p')\, D_{s0}^-(q)\, \bar{K}^0(l)]\to \Lambda (k)\,D^-(p-k)$ is 
\begin{align}\label{eq:amp1}
	\mathcal{M} &= \ii^3 \int \frac{\mathrm{d}^4l}{(2\pi)^4} \big[g_3C \bar{u}(k)\gamma_5\slashed{l}\big](\slashed{p'}+m_n)\big[g_1u(p)\big]\nonumber\\
	&\times \big[\frac{2h}{f}l\cdot v \sqrt{m_Dm_{D_{s0}}}\big]\frac{1}{{p'}^2-m_n^2}\frac{1}{q^2-m_{D_{s0}}^2}\nonumber\\
	&\times \frac{1}{l^2-m_{K}^2}\mathcal{F}^2(l^2,\,m_K^2)\,.
\end{align}
For the $\pcsstar$ as the $nD_{s1}^-$ molecular state, there are two possible amplitudes, corresponding respectively to the $J^P = 1/2^+$ and $J^P = 3/2^+$. For the former,
\begin{align}
	\mathcal{M} &= \ii^3 \int \frac{\mathrm{d}^4l}{(2\pi)^4} \big[g_3C \bar{u}(k)\gamma_5\slashed{l}\big](\slashed{p'}+m_n)\big[g_1^\prime \gamma_5 \nonumber\\
	&\times\tilde{g}_{\mu\nu}(p,m_P)\gamma^\mu u(p)\big]
	 \big[\frac{2h}{f}l\cdot v \epsilon_\alpha^\ast(p-k)\sqrt{m_{D^\ast} m_{D_{s1}}}\big]\nonumber\\
	 &\times\frac{1}{{p'}^2-m_n^2}\frac{\tilde{g}^{\nu\alpha}(q,m_{D_{s1}})}{q^2-m_{D_{s1}}^2}
	 \frac{1}{l^2-m_{K}^2}\mathcal{F}^2(l^2,\,m_K^2)\,,
\end{align}
and for the later,
\begin{align}\label{eq:amp3}
	\mathcal{M} &= \ii^3 \int \frac{\mathrm{d}^4l}{(2\pi)^4} \big[g_3C \bar{u}(k)\gamma_5\slashed{l}\big](\slashed{p'}+m_n)\big[g_1'' u_\nu(p)\big] \nonumber\\
	&\times\big[\frac{2h}{f}l\cdot v \epsilon_\alpha^\ast(p-k)\sqrt{m_D{^\ast} m_{D_{s1}}}\big]\frac{1}{{p'}^2-m_n^2}\nonumber\\
	&\times\frac{\tilde{g}^{\nu\alpha}(q,m_{D_{s1}})}{q^2-m_{D_{s1}}^2}
	\frac{1}{l^2-m_{K}^2}\mathcal{F}^2(l^2,\,m_K^2)\,.
\end{align}
Here the factor $\sqrt{m_{D^{(\ast)}}m_{D_{s0(s1)}}}$ accounts for the nonrelativistic normalization of the heavy meson fields. For the $\Sigma^{0(-)}$ production cases, the amplitudes are of the same form in Eqs. \eqref{eq:amp1}-\eqref{eq:amp3}, with varying only the corresponding meson masses.

\section{Numerical Results and Discussion}\label{sec:results}
\subsection{\texorpdfstring{$\pcs$}{} and \texorpdfstring{$\pcsstar$}{} productions}

In Ref. \cite{aaij2024a[x-}	a suspected structure were presented in the $D^-\Lambda$ invariant mass distributions with a mass of about $3.3~\mathrm{GeV}$. Consequently, we take $m_{P_{\bar{c}s}^{-}} = 3.3~\mathrm{GeV}$ in the following calculations. In this scenario, the ${P_{cs}^-}$ can still be treated as a $\bar{K}\bar{D}N$ three-body bound state, but its mass exceeds the $nD_{s0}^-$ and $D^-\Lambda(1405)$ thresholds. Thus, we determine the effective coupling by solving Eqs.~(\ref{eq:comp1}) and (\ref{eq:comp2}). The solution also relies on the width $\Gamma_{\pcs}$. Given that the $m_{P_{cs}^-}$ is tens of MeV higher than the $nD_{s0}^-$ and $D^-\Lambda(1405)$ thresholds, the phase space is significant, thereby suggesting a large decay width. To ensure a physically meaningful solution to the system of equations, we set $\Gamma_{\pcs}=200$ MeV. Consequently, we obtain the solution $|\gamma_1|\approx 11.5$ GeV and $|\gamma_2| \approx 10.7$ GeV, leading to $|g_1| \approx 3.25$.

For the mass of ${P_{\bar{c}s}^{*-}}$, we take $m_{P_{\bar{c}s}^{*-}}-m_{P_{\bar{c}s}^{-}} =m_{D_{s1}}-m_{D_{s0}}$, namely $m_{P_{\bar{c}s}^{*-}} \approx 3.44~\mathrm{GeV}$, and $\Gamma_{\pcsstar}=200$ MeV. Then for the effective coupling, we have $|g_1^\prime| \approx 1.81$ and $|g_1''| \approx 3.14$.

\begin{figure}
	\centering
	\includegraphics[width=0.95\linewidth]{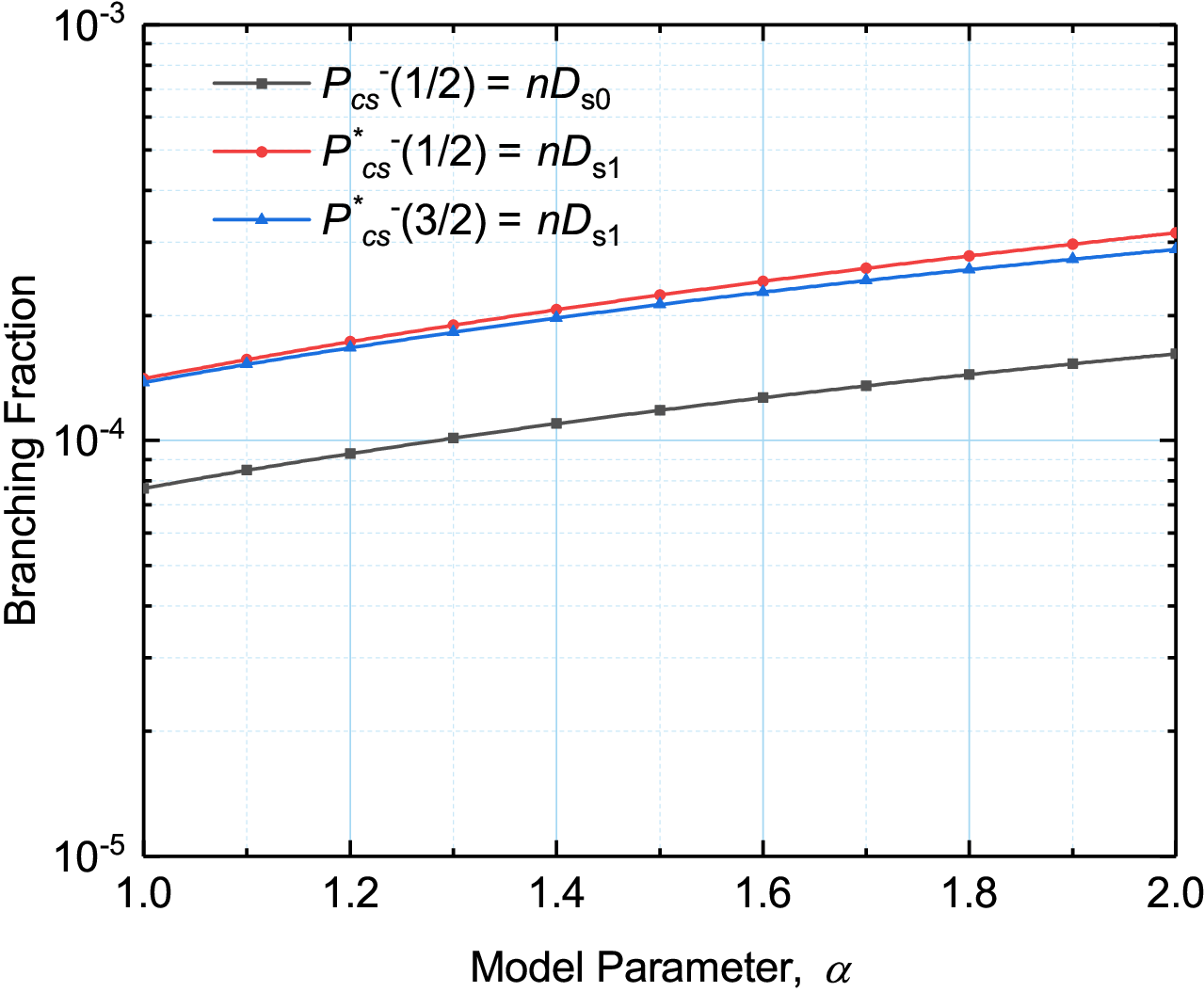}
	\caption{The $\alpha$ dependence of the branching fractions for the processes $\Lambda_b^0\to P_{\bar{c}s}^{(*)-} D^+$. According to the measurement by the LHCb collaboration \cite{aaij2024a[x-}, the mass of the $P_{\bar{c}s}^{-}(P_{\bar{c}s}^{*-})$ was taken to be $3.3(3.44)~\mathrm{GeV}$. In the calculations, $f_{D_{s0}^-}=114~\mathrm{MeV}$ and $f_{D_{s1}^-} = 194~\mathrm{MeV}$ were adopted.}
	\label{fig:bfracpcbars}
\end{figure}

Figure \ref{fig:bfracpcbars} shows the variation of the branching fractions for the $\Lambda_b^0\to P_{\bar{c}s}^{(*)-}  D^+$ with the model parameter $\alpha$ ranging from $1.0$ to $2.0$. If the $\pcs$ is the $nD_{s0}^-$ molecular state, our calculation predicts the branching fraction 
\begin{equation}
	\mathcal{B}[\Lambda_b^0\to \pcs D^+]=(0.77\sim 1.62)\times 10^{-4}\,.
\end{equation}
However, when we consider a $nD_{s1}^-$ molecule, there would be two cases, namely, the $n$ and $D_{s1}^-$ could form the $1/2^+$ or $3/2^+$ $\pcsstar$. For the former case, the branching fraction is predicted to be
\begin{equation}
	\mathcal{B}^{1/2}[\Lambda_b^0\to \pcsstar D^+]=(1.41\sim 3.16)\times 10^{-4}\,,
\end{equation}
while the branching fraction for the later case is a bit smaller:
\begin{equation}
	\mathcal{B}^{3/2}[\Lambda_b^0\to \pcsstar D^+]=(1.38\sim 2.88)\times 10^{-4}\,.
\end{equation}

As the explicit mass and width of the $\pcs$ are still experimentally unknown, we also explore several other scenarios for its production rate in $\Lambda_b^0$ decays. The numerical results are presented in Table \ref{tab:brpcsbigmass}.

\begin{table}
\caption{The branching fractions of the $\Lambda_b^0\to \pcs D^+$ for various combinations of the $\pcs$ masses (greater than the $nD_{s0}^-$ threshold) and widths. The model parameter $\alpha$ is fixed to be 1.5 in these calculations.}
\label{tab:brpcsbigmass}
\begin{ruledtabular}
\begin{tabular}{cccc}
Mass (GeV)                 & Width (MeV) & Coupling $g_1$ & Fraction($10^{-4}$) \\
\colrule
\multirow{3}{*}{3.28} & 140         & 4.17           &     2.09          \\
                      & 90          & 2.86           &     0.98          \\
                      & 40          & 0.61           &     0.045         \\
\colrule
\multirow{3}{*}{3.29} & 200         & 4.62           &    2.47           \\
                      & 150         & 3.19           &    1.18           \\
                      & 100         & 0.56           &    0.036         
\end{tabular}
\end{ruledtabular}
\end{table}

\begin{figure}
	\centering
	\includegraphics[width=0.95\linewidth]{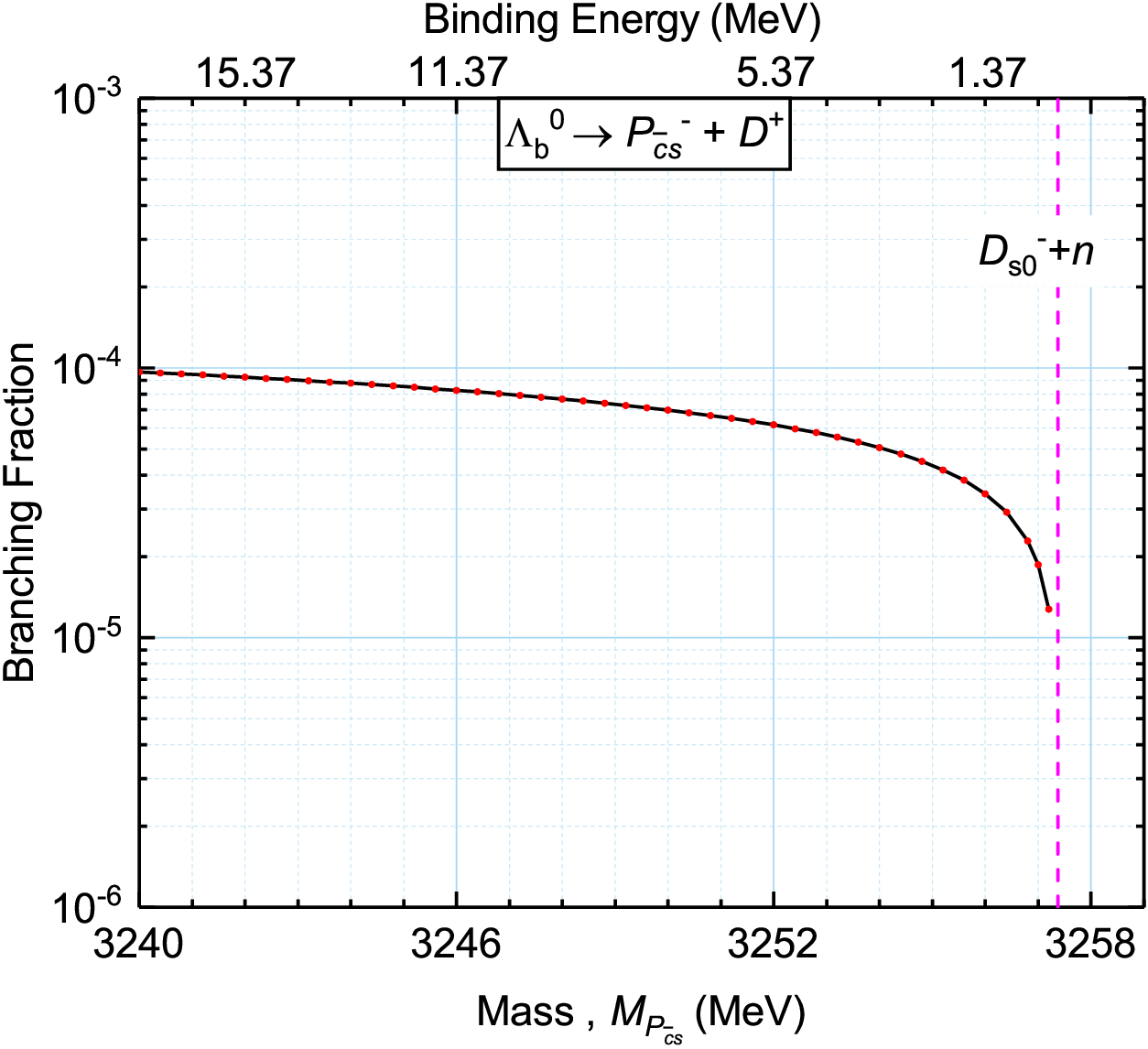}
	\caption{The branching fraction of the $\Lambda_b\to\pcs D^+ $ as a function of the $\pcs$ mass, considering the $\pcs$ as a bound state of $nD_{s0}^-$ with a binding energy of $1$--$17$ MeV. The vertical dashed line indicates the $nD_{s0}^-$ threshold. The model parameter $\alpha=1.5$.}
	\label{fig:branchvsmass}
\end{figure}

We also consider the scenario where the $\pcs$ is treated as a $nD_{s0}^-$ bound state. In this case Eq. \eqref{eq:couplingg1} is used to calculate the effective coupling $g_1$. Figure~\ref{fig:branchvsmass} shows the branching fraction of the $\Lambda_b^0\to\pcs D^+$ for different masses of the $\pcs$. As the binding energy increases from 1 MeV to 17 MeV, the branching fraction of $\Lambda_b^0\to\pcs D^+$ ranges from $10^{-5}$ to $10^{-4}$.

\subsection{\texorpdfstring{$\pcs$}{} and \texorpdfstring{$\pcsstar$}{} decays}

\begin{figure}
	\centering
	\includegraphics[width=0.95\linewidth]{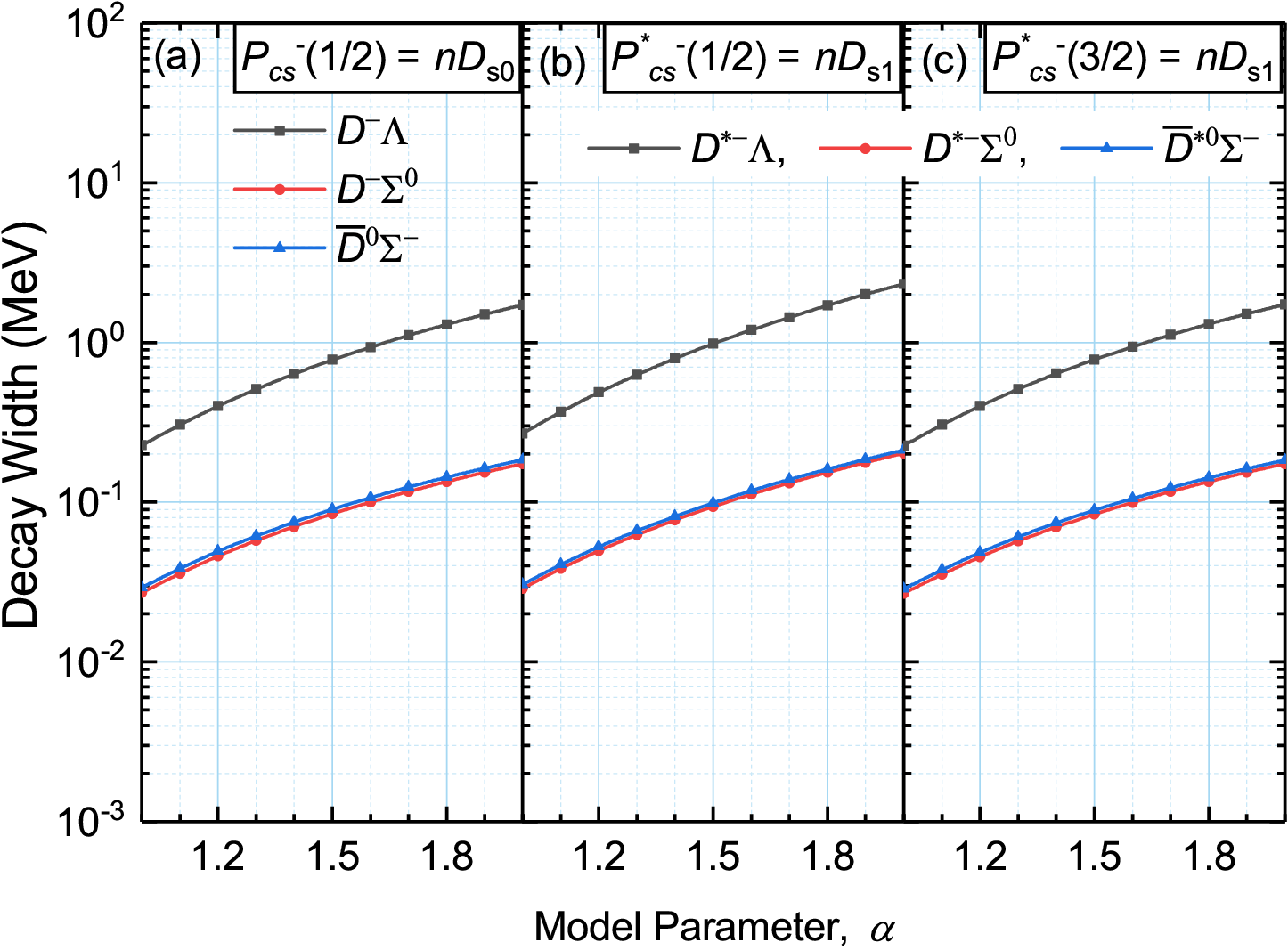}
	\caption{The partial decay widths of the $P_{\bar{c}s}^{(*)-} \to \Lambda(\Sigma) D^{(\ast)-} $ as a function of the model parameter $\alpha$. As indicated, the graph (a) stands for the $\pcs$ as the $nD_{s0}^-$ molecular state decaying into the $D^-\Lambda$, $D^-\Sigma^0$, and $\bar{D}^0\Sigma^-$; The graphs (b) and (c), respectively, describe the $\pcsstar$ as the $nD_{s1}^-$ molecular state with quantum numbers $J^P=1/2$ and $J^P=3/2$, which both go to the $D^{\ast -}\Lambda$, $D^{\ast -}\Sigma^0$, and $\bar{D}^{\ast 0}\Sigma^-$. The mass of the $P_{\bar{c}s}^{-}\ (P_{\bar{c}s}^{*-})$ was taken to be $3.3\ (3.44)~\mathrm{GeV}$.}
	\label{fig:decaywidthpcs}
\end{figure}

Finally, we calculated the partial decay widths of the $P_{\bar{c}s}^{(*)-} \to \Lambda (\Sigma^0) D^{(\ast)-}$ and $P_{\bar{c}s}^{(*)-}\to\Sigma^- \bar{D}^{(\ast)0}$ processes using the triangle loop diagrams in Fig. \ref{fig:feyndiagsdecay}. The calculated results are presented in Fig. \ref{fig:decaywidthpcs} for three interpretations of the $P_{\bar{c}s}^{(*)-}$. For the $\pcs$ as the molecule of the $nD_{s0}^-$, it would go to $D^-$ and $\Lambda(\Sigma)$; the $\pcsstar$ would decay into $D^{\ast -}$ and $\Lambda(\Sigma)$ if it is interpreted as the $nD_{s1}^-$ molecular state. The present model calculations indicate that the rates of the $\pcs$ and $\pcsstar(1/2)$ decaying into the $\Lambda D^{(\ast) -}$ are approximately equal. Moreover, the $P_{\bar{c}s}^{(\ast)-}$ exhibit the nearly same decay rates to the $\Sigma^0D^{(\ast)-}$ and $\Sigma^-\bar{D}^{(\ast)0}$. With the model parameter $\alpha$ ranging from 1.0 to 2.0, the partial widths are summarized in Table \ref{tab:decaywidthspcs}.

\begin{table}
	\caption{Partial decay widths for the processes $P_{\bar{c}s}^{(\ast)-}\to\Lambda D^{(\ast)-}$ and $P_{\bar{c}s}^{(\ast)-}\to\Sigma \bar{D}^{(\ast)}$, in units of MeV. For more information see the caption of Fig. \ref{fig:decaywidthpcs}.}
	\label{tab:decaywidthspcs}
	\begin{ruledtabular}
	\begin{tabular}{lccc}
    Final States	&$\pcs$& $\pcsstar(1/2)$ & $\pcsstar(3/2)$\\ \colrule
	$\Lambda D^{-}$ & $0.23\sim 1.72$&$\cdots$ & $\cdots$ \\
	$\Sigma^0 D^{-}$	& $0.027\sim 0.17$ &$\cdots$& $\cdots$ \\
	$\Sigma^- \bar{D}^{0}$	& $0.029\sim 0.18$  &$\cdots$& $\cdots$\\
	$\Lambda D^{\ast -}$ & $\cdots$&$0.27\sim 2.32$ & $0.23\sim 1.74$ \\
	$\Sigma^0 D^{\ast -}$	& $\cdots$ &$0.029\sim 0.20$& $0.027\sim 0.17$ \\
	$\Sigma^- \bar{D}^{\ast 0}$	& $\cdots$  &$0.031\sim 0.21$& $0.029\sim 0.18$
	\end{tabular}
	\end{ruledtabular}
\end{table}

It is seen from Fig. \ref{fig:decaywidthpcs} that the partial decay widths for all cases increase clearly in a similar manner with varying the model parameter $\alpha$. Consequently, we could define the following width ratios
\begin{subequations}
	\begin{align}
		R_1 &= \frac{\Gamma(P_{\bar{c}s} \to \Lambda \bar{D})}{\Gamma(P_{\bar{c}s} \to \Sigma \bar{D})}\,,\\
		R_2 &= \frac{\Gamma(P_{\bar{c}s}^{*}(1/2) \to \Lambda \bar{D}^{\ast})}{\Gamma(P_{\bar{c}s}^{*} (1/2)\to \Sigma \bar{D}^{\ast})}\,,\\
		R_3 &= \frac{\Gamma(P_{\bar{c}s}^{*}(3/2) \to \Lambda \bar{D}^{\ast})}{\Gamma(P_{\bar{c}s}^{*} (3/2)\to \Sigma \bar{D}^{\ast})}\,,
	\end{align}
\end{subequations}
which are expected to depend weakly on the model parameter $\alpha$. The calculated ratios are 
\begin{subequations}
	\begin{align}
		R_1 &= 4.1\sim 4.9\,,\\
		R_2 &= 4.5\sim 5.6\,,\\
		R_3 &= 4.1 \sim 5.0 \,.
	\end{align}
\end{subequations}
These ratios are insensitive to the model parameter and may be better quantities to be tested by the future experiments.

\section{Summary}\label{sec:summary}
We investigate the production of the anticharmed pentaquark states $P_{\bar{c}s}^{(*)-}$ in the $\Lambda_b^0\to P_{\bar{c}s}^{(*)-} D^+$ and their decay processes $P_{\bar{c}s}^{(*)-} \to \Lambda D^{(\ast)-}$. In the calculation, we identify the $P_{\bar{c}s}^{(*)-}$ as the $nD_{s0(s1)}^{-}$ molecule with the valence quark contents $\bar{c}sudd$, of which the mass is around $3.3~\mathrm{GeV}$ according to the LHCb experiment \cite{aaij2024a[x-}. The quantum numbers of the $\pcs$ are $J^P=1/2^+$, and for the $\pcsstar$ $J^P = 1/2^+$ or $J^P = 3/2^+$. The processes of interest occur via the triangle loops at the hadron level.

With the moderate model parameters, the computed results suggest that the branching fractions of $\Lambda_b^0\to P_{\bar{c}s}^{(*)-} D^+ $ can reach approximately $10^{-4}$.
The partial widths of the $ P_{\bar{c}s}^{(*)-} \to \Lambda D^{(\ast)-}$ range from $0.1$ to $2~\mathrm{MeV}$ with a reasonable range of model parameters. In view of the $SU(3)$ symmetry, the $P_{\bar{c}s}^{(*)-}$ would be expected to be also observed in the processes $\Lambda_b^0\to D^+\bar{D}^{(\ast)-}\Sigma$. Our model calculations reveal that the decay rates of the $P_{\bar{c}s}^{(*)-}\to \bar{D}^{(\ast)-}\Sigma $ is about one order of magnitude smaller than that of $P_{\bar{c}s}^{(*)-}\to \bar{D}^{(\ast)-}\Lambda $. It should be noted that the obtained decay widths depend on the model parameter, but their relative ratios are nearly model independent. We hope that the predictions presented here can be verified by future experiments, such as those conducted by the LHCb collaboration.

\begin{acknowledgments}\label{sec:acknowledgements}
Shi-Dong Liu thanks Qi Wu at Henan Normal University for the instruction in the calculation programs and acknowledges Tianjin University for its hospitality during his visit. This work is partly supported by the National Natural Science Foundation of China under Grant Nos. 12105153, 11975165, 12235018, 12475081, and 12075133, and by the Natural Science Foundation of Shandong Province under Grant Nos. ZR2021MA082 and ZR2022ZD26. It is also supported by Taishan Scholar Project of Shandong Province (Grant No.tsqn202103062).	
\end{acknowledgments}

\onecolumngrid
\appendix
\section{Fitting of form factors}\label{app:fff}
	\begin{figure*}[h]
	\centering
	\includegraphics[width=0.86\linewidth]{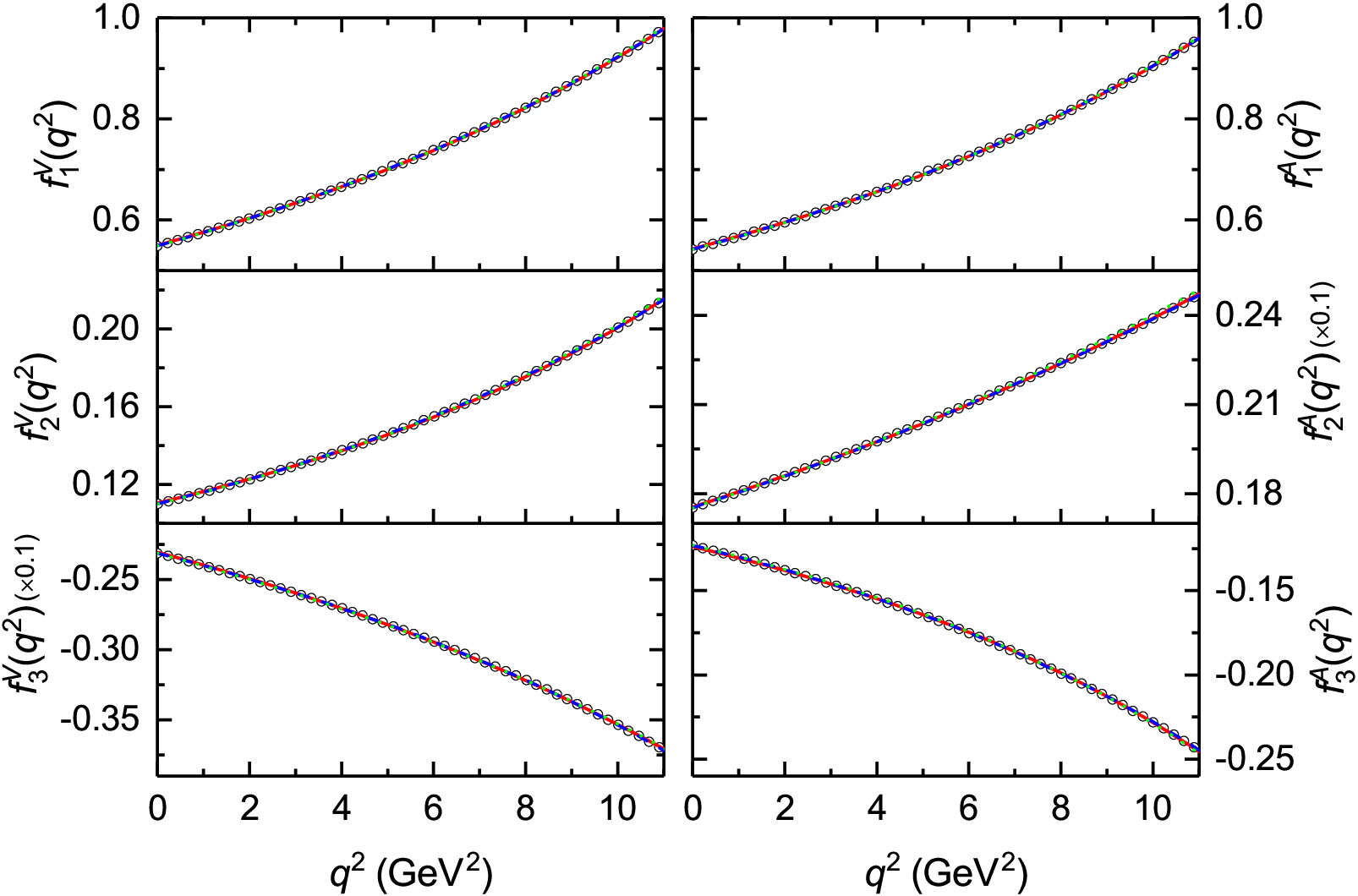}
	\caption{Fitting of the form factors in Eq. \eqref{eq:LamcLamb}. The data indicated by the open circles are obtained from Ref. \cite{gutsche2015PRD91-074001}. The red solid lines represent our re-fitting using Eq. \eqref{eq:formfactors}, the blue dashed lines are the results by Eq. \eqref{eq:fffother} with the parameters in Ref. \cite{gutsche2015PRD91-074001}, and the green dotted lines denote the results with the parameters in Ref. \cite{wu2019PRD100-114002}.}
	\label{fig:fff}
\end{figure*}

As the treatments in Refs. \cite{wu2023PRD107-054044,Wu:2021cyc,wu2019PRD100-114002}, in the calculations we usually do not use the following form of the form factor \cite{gutsche2015PRD91-074001}
	\begin{equation}\label{eq:fffother}
	    f_i (q^2) = \frac{F_i(0)}{1-as+bs^2} \quad\text{with}\quad s = \frac{q^2}{M_{\Lambda_b^0}^2}\,,
	\end{equation}
but adopt the form of Eq. \eqref{eq:formfactors} or the form below \cite{wu2023PRD107-054044,Wu:2021cyc,wu2019PRD100-114002} 
\begin{equation}\label{eq:fffotherx}
    f_i (q^2) = F_i(0) \frac{\Lambda_{i1}^2}{q^2- \Lambda_{i1}^2}\frac{\Lambda_{i2}^2}{q^2- \Lambda_{i2}^2}\,.
\end{equation}

In this appendix, we present in Fig. \ref{fig:fff} our re-fitting (red solid lines) of the form factors $f_i^{V(A)}$'s in Eq. \eqref{eq:LamcLamb} using Eq. \eqref{eq:formfactors}. The data (open circles) were obtained within the dynamical quark model by Thomas Gutsche et. al \cite{gutsche2015PRD91-074001}. For comparison, we also exhibit the results by Eqs. \eqref{eq:fffother} and \eqref{eq:fffotherx}, indicated as the blue dashed and green dotted lines, respectively. It is seen that both Eqs. \eqref{eq:formfactors} and \eqref{eq:fffotherx} can describe very well the numerical data as well as the early approximated formula \eqref{eq:fffother}.

\twocolumngrid
\bibliography{particlePhys.bib}
\end{document}